\documentclass[journal]{IEEEtran}

\ifCLASSINFOpdf
  \usepackage[pdftex]{graphicx}
  \usepackage{caption2}
  \usepackage{epstopdf}
\else

\fi

\usepackage{amsmath}
\usepackage{array}
\usepackage[colorlinks, citecolor=black, linkcolor=black, anchorcolor=black]{hyperref}
\usepackage{booktabs}
\usepackage{cite}
\usepackage{url}
\usepackage{enumerate}
\usepackage{colortbl}
\usepackage{multirow}

\definecolor{lightblue}{rgb}{.792,0.882,1}

\begin{document}

\title{Toward Better Understanding of Saliency Prediction in Augmented 360 Degree Videos}

\author{Yucheng Zhu, Xiongkuo Min, DanDan Zhu, Ke Gu, Jiantao Zhou, Guangtao Zhai,\\ Xiaokang Yang, and  Wenjun Zhang
\thanks{

Yucheng Zhu, Guangtao Zhai, Xiongkuo Min, Dandan Zhu, Xiaokang Yang and Wenjun Zhang are with the Institute of Image Communication and Network Engineering, and Artificial Intelligence Institute, Shanghai Jiao Tong University, Shanghai, 200240, China (e-mail:zyc420@sjtu.edu.cn).

Ke Gu is with the Beijing Key Laboratory of Computational Intelligence and Intelligent System, Faculty of Information Technology, Beijing University of Technology, Beijing 100124, China.

Jiantao Zhou is with the State Key Laboratory of Internet of Things for Smart City, and Department of Computer and Information Science, University of Macau, Macau, 999078, China.
}}

\markboth{ZHU \MakeLowercase{\emph{et al.}}: Toward Better Understanding of Saliency Prediction in Augmented 360 Degree Videos}%
{Shell \MakeLowercase{\textit{et al.}}: Bare Demo of IEEEtran.cls for IEEE Journals}

\maketitle

\begin{abstract}
Augmented reality (AR) overlays digital content onto the reality.
In AR system, correct and precise estimations of user’s visual fixations and head movements can enhance the quality of experience by allocating more computation resources for analysing, rendering and 3D registration on the areas of interest. However, there is inadequate research about understanding the visual exploration of users when using an AR system or modeling AR visual attention. To bridge the gap between the saliency prediction on real-world scene and on scene augmented by virtual information, we construct the ARVR saliency dataset with 12 diverse videos viewed by 20 people. The virtual reality (VR) technique is employed to simulate the real-world. Annotations of object recognition and tracking as augmented contents are blended into the omnidirectional videos.
The saliency annotations of head and eye movements for both original and augmented videos are collected and together constitute the ARVR dataset.
We also design a model which is capable of solving the saliency prediction problem in AR.
Local block images are extracted to simulate the viewport and offset the projection distortion.
Conspicuous visual cues in local viewports are extracted to constitute the spatial features. The optical flow information is estimated as the important temporal feature.
We also consider the interplay between virtual information and reality.
The composition of the augmentation information is distinguished, and the joint effects of adversarial augmentation and complementary augmentation are estimated.
We generate a graph by taking each block image as one node.
Both the visual saliency mechanism and the characteristics of viewing behaviors are considered in the computation of edge weights on the graph which are interpreted as Markov chains.
The fraction of the visual attention that is diverted to each block image is estimated through equilibrium distribution of this chain.
Extensive experiments are conducted to demonstrate the effectiveness of the proposed method.

\end{abstract}

\vspace{-0.1cm}
\vspace{-0.2cm}

\IEEEpeerreviewmaketitle

\newenvironment{rcase}
    {\left\lbrace\begin{aligned}}
    {\end{aligned}\right.}

\section{Introduction}
\IEEEPARstart{A}ugmented reality (AR) blends the computer-generated perceptual information into the real world and enhances users' visual perception of reality.
To provide meaningful information, AR must understand the context of physical world \cite{azuma1993tracking}.
Based on the content analysis and understanding, AR technology requires accurate 3D registration of virtual and real objects \cite{van2010survey}.
With the help of fast-growing computer vision technologies and advanced display technologies, AR is entering more and more application fields \cite{furht2011handbook, billinghurst2002collaborative}, e.g., by combining AR with real-time object recognition and tracking technology, users are able to have a fast understanding of surroundings.
One intuitive example is that when people are trying to find one target in a crowded room, a bit more time might be wasted if they are not familiar with the surroundings.
But if an AR system can help to analyse the scene and provide the auxiliary information, the user will quickly figure out the situation.

\begin{figure}[!t]
\centering
\includegraphics[width=3.58in]{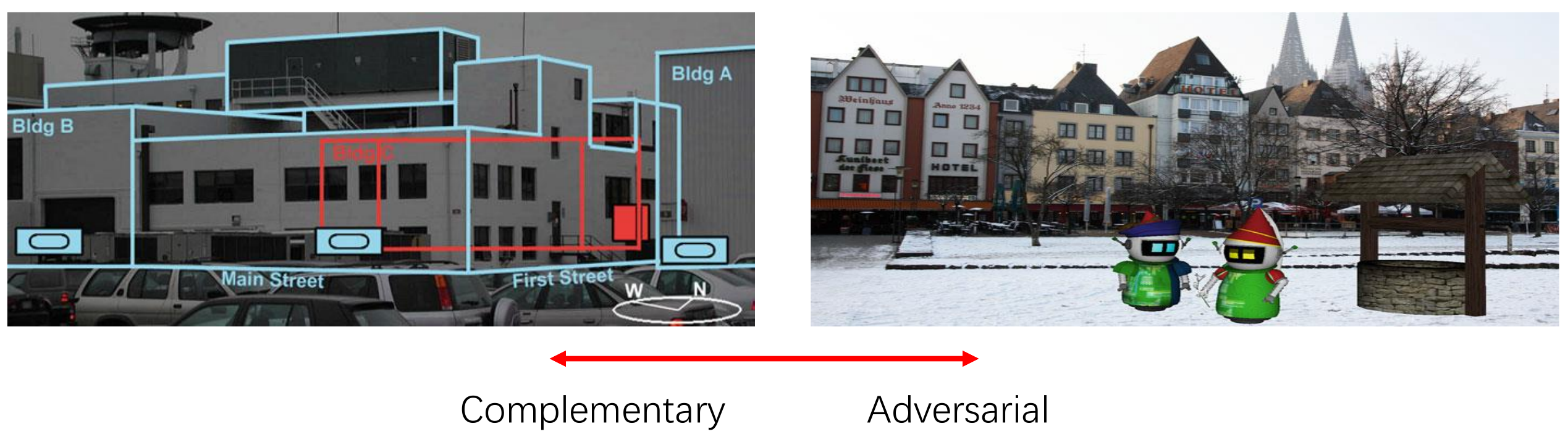}
\centering
\caption{The augmentation can be complementary or adversarial. In many cases, the augmentation is a combination of the complementary and adversarial information. The two sample images are selected from \cite{furht2011handbook}. }
\label{fig_1}
\vspace{-0.35cm}
\end{figure}

The work in \cite{PerceptualinAR} has researched the perceptual issues in AR.
Some panoramic images have the resolution of 8K, and such a large image will bring about difficulties for the storage and transmission. Besides, there are two separate views for left and right eyes, which can be used to bring stereoscopic perception, but will increase the data size at the same time.
In the past several years, some saliency prediction models have been proposed \cite{Salsurvey1}.
Visual attention has its wide applications in video compression and quality assessment \cite{yang2018saliency, paul2018efficient, yang2016principal, huynh2011importance}.
Therefore, it is meaningful to develop saliency prediction techniques to allocate processing resources more reasonably in AR system.
In bottom-up saliency prediction algorithms, low-level features are derived from the sensory information using the feed-forward model structure. In top-down models, high-level features make more explorations of how the attention is influenced by contextual cues.
And in the framework for video saliency prediction, both spatial features as well as temporal features are important.
However, the problem of saliency prediction on immersive medias is still at its infancy. Some models employ the handcrafted features and follow the strategy of heuristics \cite{Rai2017Which, STARTSEV2018, BATTISTI2018, LING2018, LEBRETON2018, zhu2018prediction}. Others seek to the advent of deep learning technique \cite{cheng2018cube, Xu_2018_CVPR}.
However, most existing methods for immersive images/videos suffer the problem that the prediction accuracy is unsatisfactory, and inadequate investigations are made about the interconnection between the instantaneous viewing behavior and long-term saliency results.
Besides, for AR videos, the virtual information is overlaid on the reality. It is straightforward that the mixture of virtual contents will alter the distribution of visual attention.
The relationship between the augmentations and the real-world perception should be considered in the prediction of visual saliency.
However, existing saliency prediction models do not distinguish the relationship. 

%
In AR experience, we can find that the focus of users for the real-world and for the virtual information can be complementary or adversarial as shown in Fig. \ref{fig_1}.
The complementary augmentation usually has a strong correlation with the surroundings.
In Fig. \ref{fig_1} (left), the augmentation provides much information about the characteristics of the surroundings.
However, when the augmentation is the adversarial type, the augmentation is usually more independent from the surroundings.
In this situation, the augmentation is usually conspicuous as shown in Fig. \ref{fig_1} (right).
In many cases, the augmentation is a combination of the complementary and adversarial information.
Complementary relationship is required in some specific scenes (e.g., driving), because in such circumstances our focus for the real environment is important. 
For some other applications like virtual advertisements, the augmented contents might be designed to dominate the visual attention.
%
However, existing saliency prediction models do not make a distinction between the two different situations, which will cause the inaccurate prediction results.
In the implementation of AR, users can view the real world by the means of video see-through or optical see-through \cite{furht2011handbook}.
For the video see-through strategy, the virtual information is overlaid on the video of the real world.
And for the optical see-through strategy, the optical elements is employed through which the real world is viewed.
Because VR can provide an immersive media experience in which viewers can get a sense of presence, we turn to employ virtual reality (VR) to simulate video see-through AR.
We can simulate the see-through videos by overlaying computer-generated digital contents onto immersive videos which are regarded as real-world scenes.

For the problem of saliency prediction for the augmented omnidirectional videos, there are some concerns should be addressed.
In many cases, the augmentation is a combination of the complementary and adversarial information.
Adversarial augmentation and complementary augmentation have different influences on the visual saliency.
We should investigate the joint effects of the complementary and adversarial information on the visual saliency.
Another problem is that during the capturing, the all 360 degrees of the scene will be recorded.
The spherical field of view is usually mapped to a flat image through different projection methods, e.g., equirectangular projection and cube mapping.
If the 360 degree image is viewed in 2D, there will be geometric distortions that are caused by the projection.
Therefore, we should mitigate the projection distortions to ensure the accuracy of 2D signal processing.
Besides, when watching immersive videos, users can have very different experience to perceive the immersive content in the first person perspective in the head mounted display (HMD).
Due to the tremendous extent of the observable content and the limited extent of field of view (FoV), the viewing behaviors including head movements (HM) and eye movements (EM) are commonplaces in immersive experience.
Therefore, it is plausible to investigate the impact of instantaneous viewing behaviors on the saliency results.
%

The amount of conveyed information of the augmentation can be used to measure the level of influence on the visual saliency.
For the complementary augmentation, the augmentation can enhance the associated surrounding areas. On the contrary, the adversarial augmentation attenuates the surroundings.
So the conveyed information of the two types of augmentations play different roles according to the augmentation type.
In other words, the level of influence is determined by the amount of conveyed information, and the type of augmentations control the choice of attenuation or enhancement.
In many cases, the augmentation is a combination of the complementary and adversarial information.
Therefore, a soft-decision strategy is a better choice to distinguish the composition of the augmentation.
The motion difference between the augmented area and the surrounding area can reflect the degree of coherence, which can be used to estimate the probabilities of augmentation types.
Based on the estimated probabilities, the expectation of influence can be calculated as the joint effects of the complementary and adversarial information on the visual saliency.
In the feature extraction, we can divide the panorama into blocks to simulate the viewport and offset the projection distortion.
Considering the impact of instantaneous viewing behaviors on the saliency results, the graph can be generated with each block as one node. Then we can formulate the accompanying adjacency matrix, of which each element indicates the relationship of adjacency from one node to the other.
Weights for different edges can be determined by incorporating the visual saliency mechanism and characteristics of instantaneous viewing behaviors. 
%
The framework of utilizing graph to tackle the problem of saliency prediction is reasonable \cite{harel2007graph}, because both the features which can be low- or high-level reflecting the mechanism of human visual system (HVS) and the characteristics of human viewing behaviors can be incorporated in the transition matrix.


\begin{figure*}[t]
\centering
\includegraphics[width=6.28in]{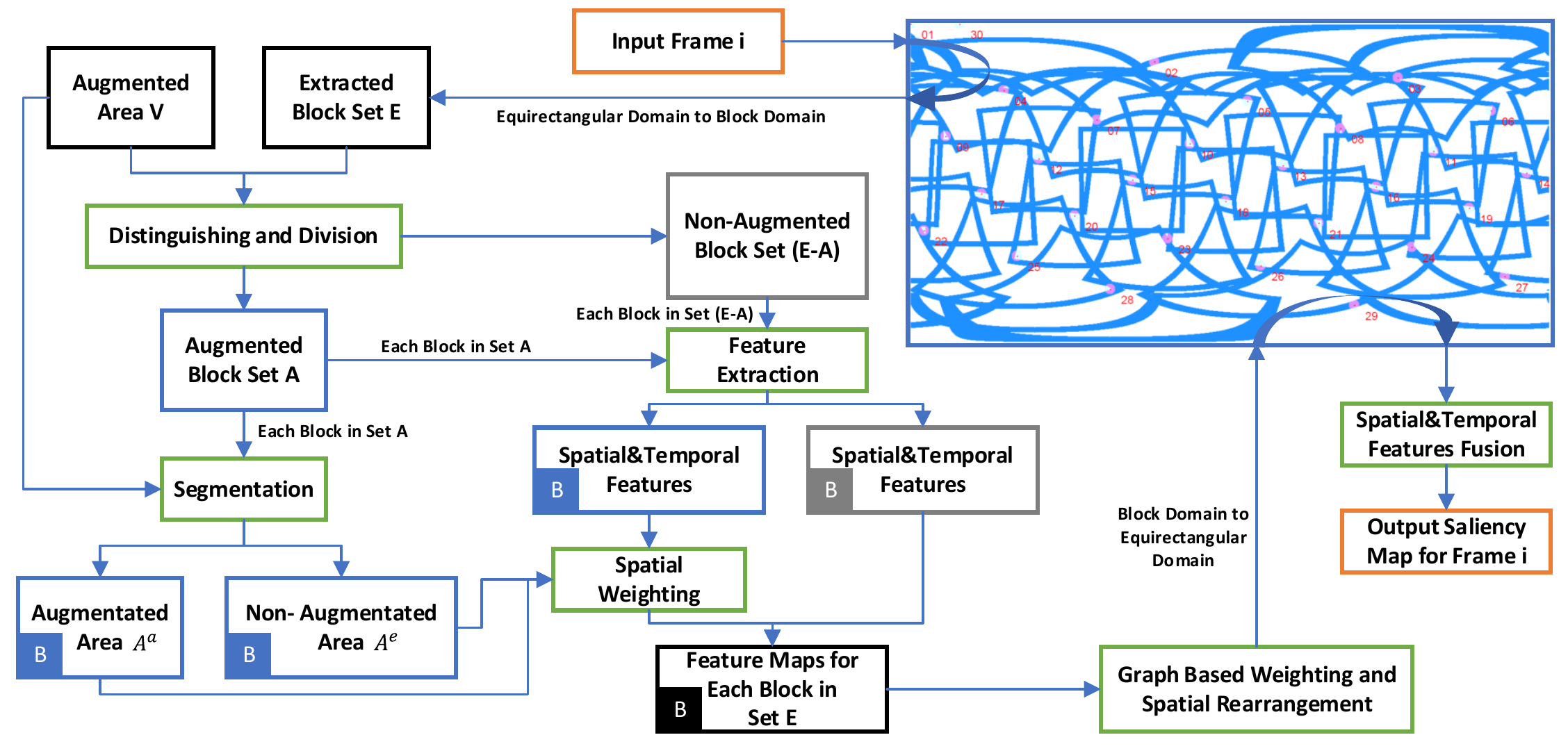}
\centering
\caption{ Framework of the proposed VBAS model. In this framework, input and output are marked with red, modules that need calculations are marked with green, results that contain the information for all block images that are extracted from the equirectangular image are marked with black, results for augmented blocks are marked with blue, and results for non-augmented blocks are marked with gray. Bottom left script ``B" indicates that the results are in block domain.
Local block images are extracted to simulate viewport and reduce the impact of distortions that are caused by projections.
The center of viewports are designed to evenly distribute on the sphere \cite{Uniformsampling, gutierrez2018toolbox}.
We distinguish the augmented block images and non-augmented block images.
On the augmented block images, the interplay between virtual information and reality is considered.
The joint effects of the complementary and adversarial information on the visual saliency are estimated, based on which the spatial weighting is conducted.
In the feature extractions, a CNN based model is designed and trained on the Large-scale scene understanding (LSUN) database \cite{LSUN} to extract conspicuous cues on local viewports.
The optical flow is calculated as the temporal feature. The FlowNet \cite{dosovitskiy2015flownet} is employed to extract motion features.
To ensure our model concentrate on significant block images, we generate the graph by taking each block image as one node.
We combine the extracted visual cues with the characteristics of viewing behaviors to define edge weights on graphs which are interpreted as Markov chains.
And different strategies are employed to fuse the spatial and temporal features.}
\label{fig_frameworkall}
\vspace{-0.25cm}
\end{figure*}

Hence, we propose a detailed methodology, named VBAS (Visual Behavior Adaptive Saliency) Model, to tackle the problem of saliency prediction on the augmented omnidirectional videos.
Our method can adapt to different visual behaviors when viewers are watching augmented immersive videos.
Specifically, local block images are extracted to simulate viewport and reduce the impact of distortions that are caused by projections.
Both spatial and temporal visual cues are extracted.
In the extraction of spatial features, we take the advantage from the adequate saliency annotations for traditional 2D images.
A convolutional neural network (CNN) based model is designed and trained on the Large-scale scene understanding (LSUN) database \cite{LSUN} to extract conspicuous cues on local viewports.
The optical flow is estimated as the temporal feature.
The FlowNet \cite{dosovitskiy2015flownet} is employed to extract motion features.
The interplay between virtual information and reality is also considered.
The joint effects of the complementary and adversarial information on the visual saliency are estimated, based on which the spatial weighting is conducted.
To ensure our model concentrate on significant block images, we generate the graph by taking each block image as one node.
We combine the extracted visual cues with the characteristics of viewing behaviors to define edge weights on graphs which are interpreted as Markov chains.
And different strategies are employed to fuse the spatial and temporal saliency.
Fig. \ref{fig_frameworkall} shows the overall framework of our method.
To validate the effectiveness of the proposed method, we construct an ARVR saliency dataset.
Considering the sense of presence in VR, we employ VR to simulate real-world environments.
We implement the blending of virtual contents by using the named bounding box to label and follow the salient objects.
We display the augmented videos in a VR HMD with an embedded eye tracker to record the head and eye movements when users are experiencing the simulated AR.
Also, the original videos are also evaluated in the VR HMD.
We generate saliency maps from head and eye movements and compare our model with some state-of-the-art models on the ARVR dataset.
Our model provides the most promising results compared with others.

The rest of this paper is organized as follows. In Section II, we summarize the main contributions of this work. In Section III, we describe the implementation details of our model to predict AR saliency. Section IV describes the construction of our ARVR dataset and the experimental evaluation results and comparisons. Finally, concluding remarks are described in Section V.

\section{Our Works}
In this paper, we make several contributions.
First, we address the differences between adversarial augmentations and complementary augmentations.
For the complementary augmentation, the augmentation can enhance the associated surrounding
areas. On the contrary, the adversarial augmentation addresses itself and attenuates the surroundings.
Second, we design the model which is capable of solving the saliency prediction problem in augmented 360 degree videos.
Local block images are extracted to simulate the viewport and offset the projection distortion.
Conspicuous visual cues in local viewports are extracted to constitute the spatial features. And the optical flow information is estimated as the important temporal feature.
We consider the interplay between virtual information and reality.
The composition of the augmentation information is distinguished, and the joint effects of adversarial augmentation and complementary augmentation are estimated.
Third, we generate the graph by taking each block image as one node.
Both the visual saliency mechanism and the characteristics of viewing behaviors are considered in the computation of edge weights on graphs which are interpreted as Markov chains.
The fraction of the visual attention that is diverted to each block image is estimated through equilibrium distribution of this chain.

Besides, we construct the ARVR dataset. The dataset is comprised of 12 original panoramic videos and 12 simulated video see-through AR sequences.
We simulate video see-through AR sequences by adding the labelled bounding box on the moving objects.
We use the ``Adobe Premiere Pro" software and spherically aware immersive video effects that can compensate for distortion in equirectangular content. The insertion of labelled bounding box is conducted manually on the sphere.
During the collection of head and eye movements data, HMD equipped with eye tracker is used.
Each of the video is saliency-annotated by 20 subjects.
On the constructed dataset, we conduct extensive experiments and demonstrate the effectiveness of the proposed method.

\section{Visual Behavior Adaptive Model Architecture}
To predict visual saliency in AR, we developed the Visual Behavior Adaptive Saliency (VBAS) prediction model based on the following considerations. Fig. \ref{fig_frameworkall} shows the overall framework of the proposed VBAS model.

\begin{enumerate}
\item \textbf{Spatial Local Saliency.} Image blocks are extracted to offset the projection distortions. CNN-based model is designed to predict the spatial saliency.
\vspace{0.1cm}
\item \textbf{Temporal Motion.} Considering that motion information is a crucial cue for visual saliency, we extract motion features as temporal saliency.
\vspace{0.1cm}
\item \textbf{Impact of Augmentation.} The impact of augmentation information is considered. The joint effects of the complementary and adversarial information on the visual saliency are estimated, based on which the spatial weighting is conducted.
\vspace{0.1cm}
\item \textbf{Viewport Saliency Fusion.} To highlight significant block maps, we design the weighting strategy by forming graphs based on block images and computing the equilibrium distribution. We consider the visual saliency mechanism and the characteristics of viewing behaviors in the calculation of edge weights on graphs which are interpreted as Markov chains.
\end{enumerate}

\begin{table}[tb]
\centering
\caption{\textsc{The Network Architecture}}
\vspace{0.2cm}
\label{tabalresnet}
\footnotesize
\begin{tabular}{l|c|c}
\toprule[.15em]
Layer Name& Output Size& Kernel, Depth, Stride \\ \midrule \midrule
Conv1 & $112\times112$ &7$\times$7, 64, stride 2 \\ \midrule
 & & $3 \times 3$ max pool, stride 2\\
Conv2\_x & $56\times56$ &
    $\left[
    \begin{array}{cc}
    1\times 1, &64\\
    3\times 3, &64\\
    1\times 1, &256
    \end{array}
    \right] \times 3$ \\ \midrule
Conv3\_x & $28\times28$ &
    $\left[
    \begin{array}{cc}
    1\times 1, &128\\
    3\times 3, &128\\
    1\times 1, &512
    \end{array}
    \right] \times 4$ \\ \midrule
Conv4\_x & $14\times14$ &
    $\left[
    \begin{array}{cc}
    1\times 1, &256\\
    3\times 3, &256\\
    1\times 1, &1024
    \end{array}
    \right] \times 6$ \\ \midrule
Conv5\_x & $7\times7$ &
    $\left[
    \begin{array}{cc}
    1\times 1, &512\\
    3\times 3, &512\\
    1\times 1, &2048
    \end{array}
    \right] \times 3$ \\ \midrule
DConv6 & $14\times 14$ & $5\times 5$, 512, stride 2 \\ \midrule
DConv7 & $28\times 28$ & $5\times 5$, 128, stride 2 \\ \midrule
DConv8 & $56\times 56$ & $5\times 5$, 32, stride 2 \\ \midrule
DConv9 & $112\times 112$ & $5\times 5$, 8, stride 2 \\ \midrule
Conv10 & $112\times 112$ & $1\times 1$, 1 \\ \bottomrule[.15em]
\end{tabular}
\end{table}

\subsection{Spatial Features Extraction}
The local saliency prediction aims to offset the pixel distortion in equirectangular projection and highlight the area of interests locally during the watching of panoramic videos.
We use the perspective projection method to map the sphere to multiple block images.
Specifically, the block image is generated by projecting the scene within a 60 degree view frustum from a given perspective centre and target position to the block image.
The 60 degree view frustum is able to cover the center vision and near peripheral vision that are most acute vision areas in eyes \cite{guastello2013human}.
The sphere centre is set as the centre of perspective.
To ensure an even distribution of target position on the sphere, we should generate the points on the sphere and make the number of points in an area proportional to the area.
Some study \cite{Uniformsampling} showed the way to achieve the even distribution on sphere, which is also employed in the toolbox to measure the performance of saliency prediction on 360 degree images \cite{gutierrez2018toolbox}
\begin{eqnarray}
\label{local1}
\gamma&=&2\pi(1-1/GR), \nonumber\\
\theta_i&=&i\gamma, \nonumber\\
\varphi_i&=&\arccos(1-2i/N), \nonumber\\
\textbf{Q}&=&\{(\varphi_i, \theta_i), i = 1... N\},
\end{eqnarray}
where N is the total number of target points, $GR = \frac{1+\sqrt{5}}{2}$, $\varphi_i$ and $\theta_i$ are latitude and longitude respectively, $\textbf{Q}$ is the set of evenly distributed target points. We can extract block images on the panorama from different target positions
\begin{eqnarray}
\label{block_extraction}
&\textbf{E} = \{E|E = \text{Extract}(I_{pano}, \varphi, \theta), (\varphi, \theta)\in \textbf{Q}\},
\end{eqnarray}
where $I_{pano}$ is the panorama. $\textbf{E}$ is the set containing extracted block images.

We design the CNN-based model to predict the spatial local saliency.
The saliency prediction model takes the block images (224$\times$224$\times$3) as input and outputs the predicted saliency maps.
The backbone of the saliency prediction model is the Resnet50 \cite{he2016deep} network, pre-trained on ImageNet \cite{krizhevsky2012imagenet}. We remove the average-pooling layer and fully-connected layer in Resnet.
To upscale the feature map to the dimension of 112$\times$112, four transpose convolution layers are employed to conduct the deconvolution.
And the final convolution layer is employed to produce the saliency map.
Table \ref{tabalresnet} shows more details.

\vspace{-0.2cm}
\subsection{Temporal Features Extraction}
Motion information is a crucial cue for visual saliency, so motion features are extracted as temporal saliency.
In the literature, the FlowNet \cite{dosovitskiy2015flownet} has the outstanding motion describing ability.
Considering that visual attention is often attracted by motion, we take the extracted motion as temporal saliency.
FlowNet is employed to detect motion on each block image in $\textbf{E}$ between frames.
To reduce the impact of noise, in our processing, we filter the magnitude of FlowNet's results with Gaussian.

To integrate the spatial and temporal saliency maps, we can employ the following feature fusion methods \cite{min2016fixation}.
\begin{itemize}
\item Product after normalization:
\begin{eqnarray}
\label{pn}
f_{1}(S_s,S_t) = \mathcal{N}\left(\prod_i \mathcal{N}(S_i) \right),
\end{eqnarray}
\item Max after normalization:
\begin{eqnarray}
\label{mn}
f_{2}(S_s,S_t) = \mathcal{N}\left(\max_i \mathcal{N}(S_i) \right),
\end{eqnarray}
\item Summation after normalization:
\begin{eqnarray}
\label{sn}
f_{3}(S_s,S_t) = \mathcal{N}\left(\sum_i \mathcal{N}(S_i) \right),
\end{eqnarray}
\end {itemize}
where in Eqs. (\ref{pn}), (\ref{mn}), (\ref{sn}), $i\in \{s,t\}$ which represents spatial and temporal domain respectively, and $\mathcal{N}$ is a normalization operator which normalizes the saliency map $S$ linearly to the dynamic range $[0,1]$.
A pixel-wise manner is adopted in the summation, product and max operations in Eqs. (\ref{pn}), (\ref{mn}), (\ref{sn}).

\subsection{Spatial Weighting on Augmented Blocks}
The complementary augmentation bears a strong correlation with the surroundings. So, complementary augmentation can highlight its associated target.
However, when the augmentation is adversarial, the virtual object is usually independent from the surroundings. So, the adversarial augmentation is usually conspicuous.
%
In many cases, the augmentation is a combination of the complementary and adversarial information.
Therefore, it is plausible to calculate the amount of complementary and adversarial information in the augmentations and approximate their joint effects on the visual saliency distribution.
The amount of conveyed information of the augmentation can be used to measure the level of influence on the visual saliency.
For the complementary augmentation, the augmentation can enhance the associated surrounding areas.
On the contrary, the adversarial augmentation attenuates the surroundings.
So the conveyed information of the two types of augmentations play different roles according to the augmentation type.
In other words, the level of influence is determined by the amount of conveyed information, and the type of augmentations control the choice of attenuation or enhancement.
Because in many cases, the augmentation is a combination of the complementary and adversarial information, the soft-decision strategy can perform better.
So we can estimate the probabilities for the two types of augmentations.
The motion difference between the augmented area and the surrounding area can reflect the degree of coherence, which can be used to estimate the probabilities of augmentation types.

Specifically, among these block images, we search the ``augmented block images" that overlap with the augmented area and satisfy that the overlap ratio is greater than 0.3. The suitable block images constitute the set $\textbf{A}$
\begin{eqnarray}
\label{block1}
&\textbf{A} = \{A|A \in \textbf{E},\frac{A\cap V}{A}>0.3\},
\end{eqnarray}
where $V$ represents the occlusion area where laid on the augmentation information, and $\textbf{A}$ consists of the augmented block images that are extracted on the panorama from different target positions in set $\textbf{E}$, and satisfy that the overlap ratio is greater than 0.3.
Each augmented block image can be further segmented on the pixel level
\begin{eqnarray}
\label{pixel1}
A^a = A\cap V, \nonumber\\
A^e = A- V,
\end{eqnarray}
where $A^a$ contains the pixels that belong to augmentations, $A^e$ contains the pixels that belong to the reality environment.
We calculate the entropy on the $A^a$ to estimate the information that is conveyed by the augmented area
\begin{eqnarray}
\label{entropy_cal}
H(A^a) = \sum_{m} -p_m\log_2(p_m)
\end{eqnarray}
where $M$ is the number of gray levels and $p_m$ is the probability associated with gray scale $m$.
We enhance the $A^a$ when it is adversarial augmentation. Otherwise, we enhance the $A^e$ when it is complementary type.
There are two augmentation types that are complementary ($c_1$) and adversarial ($c_2$). So, the augmentation type (r) has the sample space $\{c_1,c_2\}$.
In the estimation of the joint effects of the two types of augmentations, the soft-decision strategy is designed by
\begin{equation}
\label{augmentation_w2}
  \begin{rcase}
    w^a &= E_{r\sim p(r)}(2\cdot \text{Sigmoid}(K(r)\cdot H(A^a))),          \\
    w^e &= 2-w^a, \,
  \end{rcase}
\end{equation}
where $w^a$ is the weight for $A^a$, $w^e$ is the weight for $A^e$, $r\in\{c_1,c_2\}$, $E_{r\sim p(r)}$ is the expectation of the sigmoid function based on the probability $p(r)$. In Eq. (\ref{augmentation_w2}), the $K(r)$ controls the choice of attenuation or enhancement according to the augmentation type
\begin{eqnarray}
\label{direction}
K(r) &=& \left\{ \begin{array}{cc}
-1, & \text{if} \quad r=c_1\\
1, & \text{if} \quad r=c_2
\end{array} \right.
\end{eqnarray}
$K(r)$ is designed that when the augmentation is complementary, the augmented information will divert visual attention to the surrounding associated areas, otherwise, the adversarial augmentation itself will attract visual attention.
With the calculated spatial weights, we can conduct the spatial weighting by
\begin{eqnarray}
f_4(S(x,y)) = \left\{ \begin{array}{cc}
w^a \cdot S(x,y), & \text{if }(x,y)\in A^a\\
w^e \cdot S(x,y), & \text{if }(x,y)\in A^e
\end{array} \right.
\label{weighted}
\end{eqnarray}
where $S(x,y)$ is the feature value at pixel $(x,y)$.

To determine the probability $p(r)$ in Eq. (\ref{augmentation_w2}), we assume that the motion difference between the augmented area and the surrounding area can reflect the degree of coherence.
We calculate the mean of pixel flows on the given area
\begin{eqnarray}
\label{flow_cal1}
\overline{\textbf{f}^i}_t = \frac{1}{M_i}\sum_{(x,y)\in A^i} \textbf{f}_{x,y,t}
\end{eqnarray}
where $i\in\{a,e\}$ that is defined in Eq. (\ref{pixel1}), and $\textbf{f}_{x,y,t} = (u(x,y,t), v(x,y,t))$ is the flow of a pixel $(x,y)$ at time t, $M_i$ is number of pixels in $A^i$. Then we calculate the difference of the mean flow on the $A^a$ and $A^e$
\begin{eqnarray}
\label{flow_cal2}
\triangle \overline{f}_t = \|\overline{\textbf{f}^a}_t - \overline{\textbf{f}^e}_t\|_2.
\end{eqnarray}
The beliefs about the augmentation type before the prediction is considered as the prior $\pi$.
The probability of the augmentation type conditioned on the prior knowledge is calculated as
\begin{equation}
\label{prob}
  \begin{rcase}
p(c_1|\pi = c_1) &= G( \tanh(\triangle \overline{f});\mu_{c_1}, \sigma_{c_1}), \\
p(c_2|\pi = c_1) &= 1-p(c_1),\\
p(c_2|\pi = c_2) &= G( \tanh(\triangle \overline{f});\mu_{c_2}, \sigma_{c_2}), \\
p(c_1|\pi = c_2) &= 1-p(c_2), \\
  \end{rcase}
\end{equation}
where $G$ is the Gaussian, $\pi\in\{c_1,c_2\}$. $\tanh(*)$ maps the $\triangle \overline{f}$ to the range of $[0,1)$. The parameters are set as $\mu_{c_1} = 0$, $\sigma_{c_1} = 0.85$, $\mu_{c_2} = 1$, $\sigma_{c_2} = 0.85$.
Then we can use Eq. (\ref{prob}) to calculate probability $p(r)$ and use Eq. (\ref{augmentation_w2}) to conduct the spatial weighting.

\subsection{Graph based Weighting and Spatial Rearrangement}
After feature extractions in spatial and temporal domain, and spatial weighting for augmented block images, we get feature maps for each block image.
The spatial rearrangement is required to transform the block images to the equirectangular image.
If uniform weights are assigned to blocks, the resulting combined saliency map will be too uniformly distributed and uninformative.
To solve this problem, considering the visual saliency mechanism and characteristics of visual behavior, we design the graph based weighting for the spatial rearrangement.
For traditional 2D images, eye movements are nearly sufficient to view the image.
Our eyes have a limited acute vision angle of about 15 degrees in the foveal area.
To see areas accurately, we can move eyes to direct the foveal vision to the target.
Through eye movements, we can perceive most targets in front of us.
However, to explore the environment of 360 degrees, we need to change the orientation of our head.
So we should consider the characteristics of instantaneous viewing behavior when viewers are watching 360 degree videos.

We form the fully-connected directed graph $G$ by connecting every node in the set $\textbf{Q}$ that is defined in Eq. (\ref{local1}) with other $N-1$ nodes.
The weights of the outbound edges of each node can be normalized to 1.
Then we can define a Markov chain on $G$ by taking the nodes as the state, and edge weights as transition probabilities.
The fraction of the visual attention that is diverted to each node can be estimated through equilibrium distribution of this chain.
The visual saliency mechanism as well as the instantaneous visual behavior are considered in the determination of the transition probability
\begin{eqnarray}
\label{markovweight}
w(q_i,q_j) = \left\{ \begin{array}{cc}
sin(\varphi_j)\cdot \frac{\mu(S_j)}{\sum_{q_k\in \textbf{Q}_{q_i}}\mu(S_k)}\cdot F(g(q_i, q_j)) \quad \\ \text{if}\ q_j \in \textbf{Q}_{q_i},\\
sin(\varphi_j)\cdot \frac{\mu(S_i)}{\sum_{q_k\in \textbf{Q}_{q_i}}\mu(S_k)}\cdot F(g(q_i, q_j)) \quad \\ \text{if}\ q_j \notin \textbf{Q}_{q_i}
\end{array} \right.
\end{eqnarray}
where the weight of the edge from start node $p_i$ to destination node $p_j$ is proportional to the first term $sin(\varphi_j)$ that is the equator bias of the node $p_j$.
The equator bias is an assumption that viewers are more likely to hold their heads erectly when they are using HMDs.
Because the upright head position is the natural position that makes people feel comfortable.
Many elements in life take this critical consideration for ergonomic design.
So, the probability of the situation when our heads are tilted forwards or backwards is much lower than that when our heads are in upright position.

In the second term in Eq. (\ref{markovweight}), we define the set $\textbf{Q}_{q_i}$ that consists of the nodes in the field of view (FoV) centered at the node $q_i$
\begin{eqnarray}
\label{viewport_set}
\textbf{Q}_{q_i} = \{q|q\in \textbf{Q}, |\varphi-\varphi_i|<\varphi_\triangle, |\theta - \theta_i|<\theta_\Delta \},
\end{eqnarray}
where $q = (\varphi, \theta)$, $q_i = (\varphi_i, \theta_i)$. The $\varphi_\triangle$ and $\theta_\Delta$ are the half width and half height of the viewport, and $q_i$ locates at the center of the viewport.
In this paper, the data collection is conducted with the video see-through equipment with FoV of 110 degrees.
Therefore, the $\varphi_\triangle$ and $\theta_\Delta$ are set as $0.3\pi$ accordingly.
If the AR is implemented with optical see-through equipment, the parameters can be set according to the portion of space in which objects are visible at the same moment during steady fixation of gaze in one direction.

In the second term in Eq. (\ref{markovweight}), there are two cases to consider.
In the first case, the destination node is within the FoV that is centered at the start node $q_i$.
Although our eyes can only see details in the center of gaze clearly and sharply, the areas outside the central vision can also receive the light \cite{brandt1973differential}.
Therefore, eyes can perceive the content in the area that is located at $q_j$ when the fixation of gaze is in $q_i$, and more details of the area in $q_j$ will be distinguished when the gaze is shifted from $q_i$ to $q_j$.
To estimate the ability of $q_j$ to attract attention from the whole visible FoV, we calculate the ratio of the local saliency to the accumulated saliency in the FoV.
We set the numerator as $\mu(S)$ to calculate the pixels' mean value on the block feature map $S$, and set the denominator to calculate the accumulated feature value in the FoV.
In the second case, the destination node is outside the FoV that is centered at the start node $q_i$, which means that there is no perception of the area in $q_j$ when the fixation of gaze is in $q_i$.
In this case, the visual system rejects the candidates in the FoV and selects $q_j$ that is outside the FoV.
Since the content in $q_j$ is invisible, we estimate the probability that reject the candidates by $\frac{\mu(S_i)}{\sum_{q_k\in \textbf{Q}_{q_i}}\mu(S_k)}$, where $S_i$ is the block feature map that is centered at $q_i$.

In the third term in Eq. (\ref{markovweight}), we estimate the influence of the gaze shift distance.
%
According to statistics, when the size of destination area remains unchanged, the probability that the fixation locates within the destination area is inverse proportional to the active area.
Therefore, the probability that the gaze is directed within the destination area is inverse proportional to the distance of the gaze shift.
So, $F(\cdot)$ is chosen to impose a penalty on the distance.
Since image pixels are located on sphere, the distance between two points is measured along the surface of the sphere. So the geodesic distance $g(q_i, q_j)$ is calculated.
As suggested by \cite{harel2007graph}, a Gaussian function is utilized to generate the weight for distance. We set $F(x) = exp(\frac{-x^2}{2\sigma^2})$, and set the parameter $\sigma$ to approximately $0.25\pi$ ($45^{\circ}$) which is derived from half the maximum head turning angle ($60^\circ$) plus the best eye rotation angle ($15^\circ$) \cite{guastello2013human}.



The transition matrix $\textbf{W}$ can be formed by determining all the weights of directed edges according to Eq. (\ref{markovweight}).
We can find that all components in the transition matrix are strictly positive.
The principal eigenvector of the matrix can be computed as the equilibrium distribution.
According to Perron-Frobenius theorem \cite{chang2008perron}, a real square matrix with positive entries has a unique largest real eigenvalue, and there is the associated principal eigenvector, all of whose elements are positive.
Therefore, we can calculate the principal eigenvector $\alpha$ of the transition matrix $\textbf{W}$. Each element of $\alpha$ is the positive weight for the corresponding block map.
Then the spatial rearrangement can be conducted to transform the block feature maps to euqirectangular format
\begin{eqnarray}
\label{viewport_fusion}
f_5(\alpha, S) = \sum_{E_i\in \textbf{E}}\textbf{R}(\alpha_i \cdot S_{E_i}),
\end{eqnarray}
where $\textbf{R}$ consists of two steps that are the inverse operation of the `Extract' in Eq. (\ref{block_extraction}) followed by a Gaussian filtering to offset the discontinuity in the equirectangular domain. $\textbf{E}$ is the set containing extracted block images. $S_{E_i}$ is the block feature map for the $i_{th}$ block image.

\section{Experiments and Results}

\subsection{Subjective Experiment}
\subsubsection{Video Stimuli}
\begin{figure}[!t]
\centering
\includegraphics[width=3.55in]{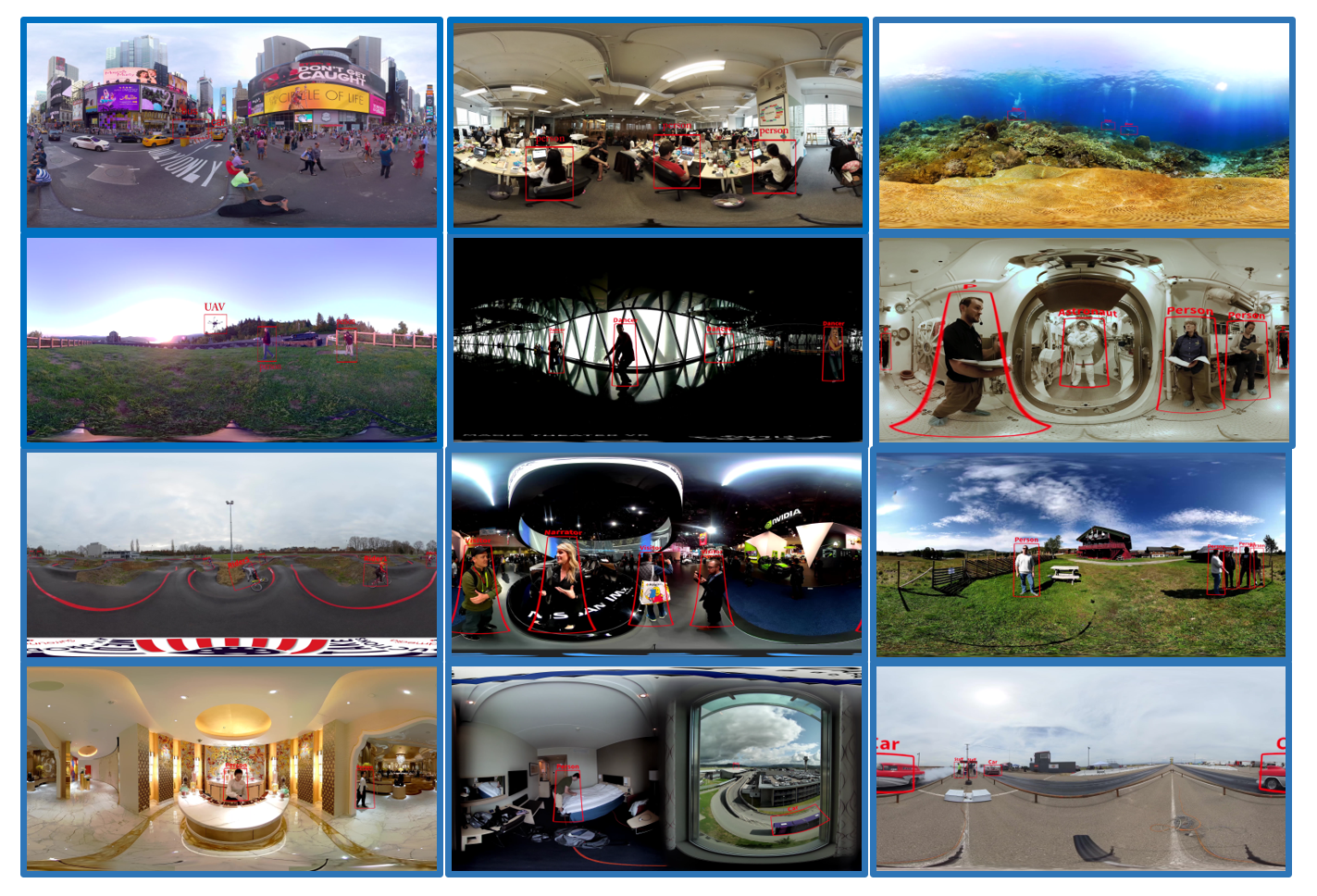}
\centering
\caption{Some example frames that are extracted from the augmented omnidirectional videos (videos annotated with bounding box) in our dataset.}
\label{fig_examples}
\vspace{-0.1cm}
\end{figure}

Considering the sense of presence in VR, we employ VR to simulate real-world environments.
In total, 12 different videos were used in the experiment.
The used videos were chosen among omnidirectional videos from YouTube.
Table \ref{table:video} provides the YouTube ID to the original omnidirectional videos hosted in YouTube.
Besides, the duration, frame rate and resolution of each video are also provided.
The selection of omnidirectional videos was carried out with the objective of covering various characteristics.
Specifically, amount of foreground objects, shooting environments, amount of motion information were considered.
The video selection resulted in a minimum of 5 videos within each one of the following groups: Indoor, Outdoor, People, Rich motion, Rare motion.
Example frame images of the videos in our dataset are shown in Fig. \ref{fig_examples}.
And all of the videos are in equirectangular format.

We implement the blending of virtual contents by using the named bounding box to label and follow the  conspicuous foreground objects.
The bounding box is designed to follow the objects to simulate the 3D registration in AR.
To ensure the quality, we use software called the ``Adobe Premiere Pro'' and conduct the annotations manually on the sphere frame by frame.
The appearance of the bounding box is shown in Fig. \ref{fig_examples}.

\begin{table}[!t]
	\centering
\tiny
\caption{\textsc{The Derivation of the Videos and the Information of each Video Clip.}}
\vspace{0.2cm}
\label{table:video}
	\begin{tabular}{c c c c c c}
		\hline
		 Stimulus name& YouTube ID &Duration(s) &Frame rate &Resolution \\
		\hline
Bar counter & HSSveMXedU4 & 38 & 24 & 4096$\times$2048 \\
Auto show & kbkSLdB5MWI & 30 & 24 & 4096$\times$2048 \\
Underwater & TkmYAjh7oQ0 & 18 & 24 & 4096$\times$2048 \\
Street dance & xZA9fw1ja1s & 30 & 24 & 4096$\times$2048 \\
Lawn & j3N4WI4wgyc & 24 & 30 & 4096$\times$2048 \\
Airport & HyoI\_VCalN8 & 30 & 30 & 4096$\times$2048 \\
Car race & Q2BtoxRtpjw & 30 & 30 & 4096$\times$2048 \\
Spacesuit Training & crt1XNocRKc & 30 & 30 & 4096$\times$2048 \\
Grassland & WYnQP8mZN9k & 30 & 30 & 3840$\times$1920 \\
Bicycle train & JT5GRWQMpy4 & 20 & 30 & 4096$\times$2048 \\
Office & WHSFeVbNzuc & 30 & 30 & 3840$\times$1920 \\
Time square & 6Yc-SXEFveo & 30 & 30 & 3840$\times$1920 \\
		\hline
	\end{tabular}
\end{table}

\begin{figure*}[t]
\centering
\includegraphics[width=5.50in]{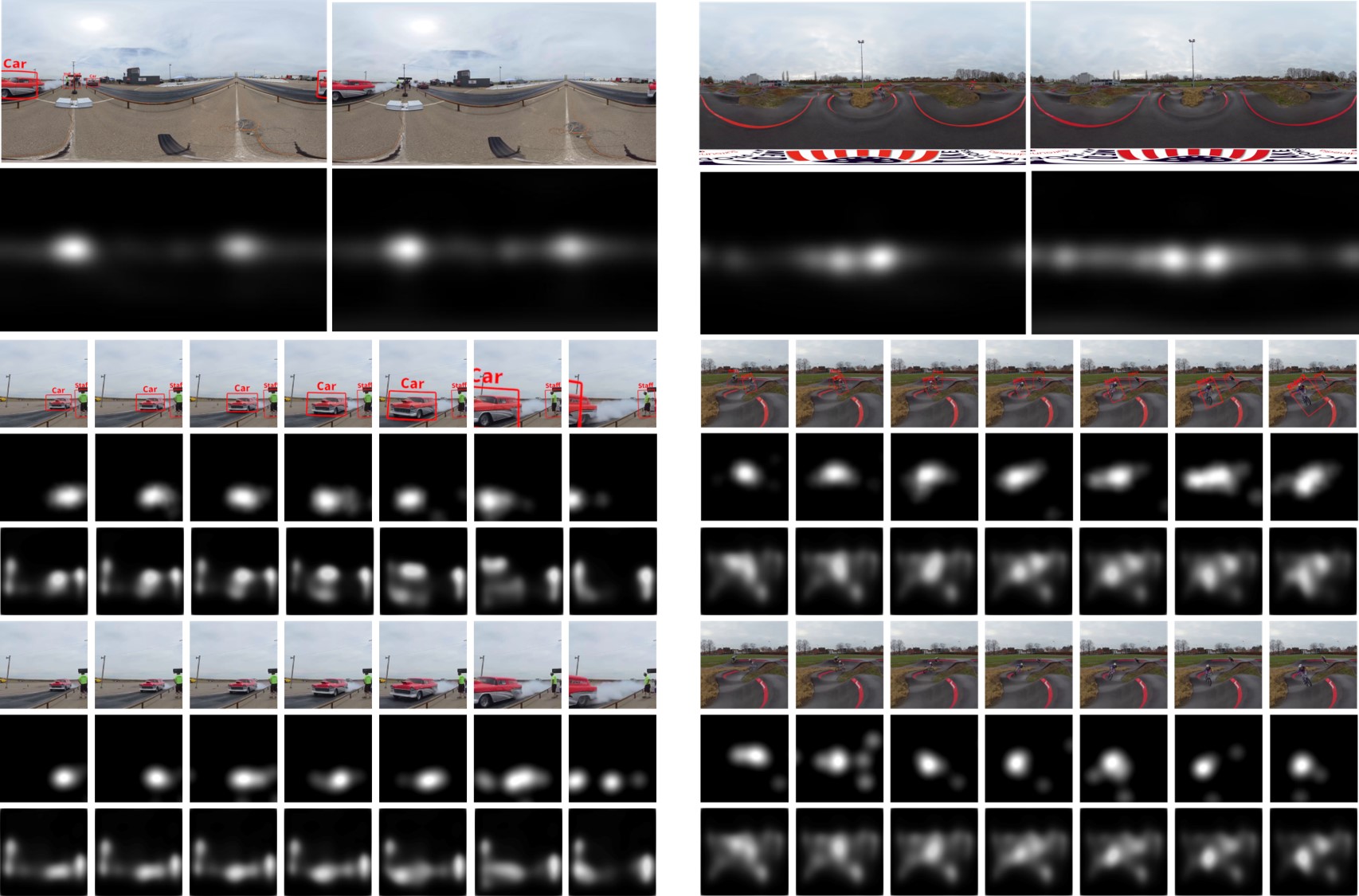}
\centering
\vspace{0.2cm}
\caption{Comparisons between the saliency maps. In each row, from top to down are the original and the augmented example frames to describe the video, and the packed saliency map for the video; the local image patch from the augmented frames, the corresponding ground-truth saliency map, and the local saliency map generated by SalGAN \cite{Pan_2017_SalGAN}; the local image patch from original frames, the corresponding ground-truth saliency map, and the local saliency map generated by SalGAN \cite{Pan_2017_SalGAN}.}
\label{fig_show}
\end{figure*}

\begin{figure*}[t]
\centering
\includegraphics[width=4.78in]{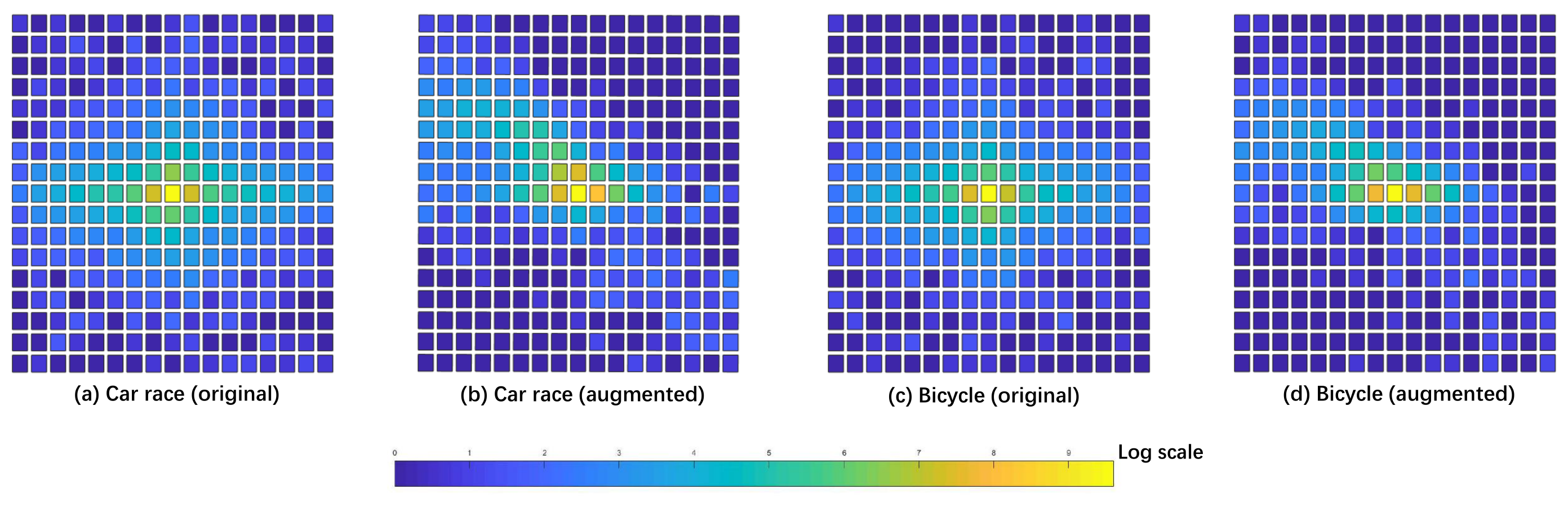}
\centering
\caption{The velocity with different directions and magnitudes are calculated and represented by the heat map.
Different colors indicate the number of end points of the velocity vector from origin in different areas. The magnitude is limited within the square with width 20, of which the unit is rad/s. The origin is set at the center of the map, and the direction is set to match the movement of human eyes.}
\label{fig_velocity}
\vspace{-0.25cm}
\end{figure*}


\subsubsection{Subjective Experiment Execution}
The `HTC VIVE PRO EYE' was employed in our experiments \cite{viveproeye}. This type of HMD can track eye movements precisely and has a wide field of view of 110 degrees, a refresh rate of 90Hz and a high resolution of $1440\times1600$ per eye. Besides, the headphones of the HMD can be used to play the audio.
The data collection software was developed based on the platform named ``Unity", running on the high-performance computer equipped with the ``1080Ti" graphics card.

Twenty subjects participated in the subjective experiment.
There were 2 females and 18 males, whose ages ranged between 20 and 30.
All of them reported normal or corrected-to-normal vision.
And the majority of the participants were familiar with the VR equipment and had the VR experience more than 5 times.
During the experiment, for the convenience of exploration, viewers were seated on a swivel chair.
The swivel chair was fixed in the ground and all subjects were only allowed to move eyes and head and rotate the swivel chair.
Carefully calibration of the eye tracking module was conducted for each subject before the recording of head and eye movements.
The training and explanation of the procedure were also provided prior the test.
The omnidirectional videos were displayed in a sequence with fixed order. 5 seconds of black screen was inserted between each two videos as a rest to make viewers be ready for the next video.
All subjects were asked to look around in this step to get natural-viewing visual attention data.
The duration of the experiment for each participant lasted less than 20 minutes, avoiding the fatigue of observers.
We collect the HM and EM data on videos with annotations as well as videos without annotations.
The data collection on the two video sets are undertook separately.
We separate the data recording for AR set and VR set with an interval of 20 days to eliminate the influence of twice experience.


\subsubsection{Gaze Data Processing}
The raw data from the head gyroscope sensor includes the 3 Euclidean rotation angles: pitch, yaw and roll.
The data of the gaze point is captured and represented in the 3D Cartesian coordinate system.
For the convenience of calculating on the sphere as well as on the 2D plane, we convert the captured data to the geographic coordinate system. All data is projected in latitude/longitude.

Saccade is the very fast eye movement between two fixations.
The sampling rate of eye tracker is 120Hz, so many recordings during the saccade is involved.
During the saccade, brain barely process the visual input. Therefore, saccade does not reflect the attention of viewers and it is better to filter out the eye movement data during the saccade \cite{nguyen2019saliency}.
However, there also exists the smooth eye pursuit in the video watching, which is made when the eye is following a moving object and is much similar with saccade but is more informative \cite{RAYNER19953}.
So we should preserve the data of the smooth eye pursuit.

To offset the stretching distortion near the poles, for each viewer, we calculate the geodesic distance between adjacent two recorded fixations.
The filtering of the data between the saccade is conducted as
\begin{eqnarray}
\label{dataprocess}
\textbf{X'} = \{x| \mu(\textbf{X})-3\sigma(\textbf{X})\leq x \leq \mu(\textbf{X})+3\sigma(\textbf{X}), x \in \textbf{X} \},
\end{eqnarray}
where $\textbf{X}$ is the set of geodesic distances of one subject during the interval of two frames.
We calculate the mean and standard deviation of the geodesic distances in $\textbf{X}$.
We remove the eye movements that are out of the range and form $\textbf{X'}$ as the gaze data of each viewer on each frame.
The same operations are conducted to filter the head movements data.

After the determination of fixations, we can generate the fixation map through aggregating all subjects' fixation points at a given time stamp.
By aggregating all subjects' data of eye movements, the fixation map for each frame is generated.
It is crucial to determine continuous region of interest.
Applying a Gaussian filter on the fixation map can tackle this problem \cite{mit-saliency-benchmark}.
The saliency level of an area is determined based on the density of fixations.
Besides, Gaussian filtering can reduce the impact of noise, calibration error, and fall-off in visual acuity outside the foveal
region \cite{john2019benchmark}.
Because the employed equirectangular projection has stretching distortions near poles, the Gaussian filtering is conducted on the sphere to offset the projection distortion of equirectangular map.
%

\subsubsection{Dataset Analytics}
Our dataset consists of the head and eye movements data, based on which we generate saliency maps for both head and eye movements on each frame image of the stimulus.
%
It is worth noting that for the adversarial augmentations, the augmentations are designed to dominate users' visual attention.
On the contrary, the complementary augmentations are designed for assistance.
With the established dataset, we can investigate how the augmentations influence the viewing behavior and analyse the differences between visual attention distributions when viewers are experiencing the AR or VR techniques.


To analyse the difference in the distribution of visual attention between AR and VR, we choose two videos, wherein
the ``car race'' is outdoor scene and contains concentrated motion and salient objects,
the ``bicycle training'' contains multiple moving salient objects.
In Fig. \ref{fig_show}, the ``packed" saliency map is calculated as Eq. (\ref{packed}) to estimate the overall distribution of visual attention
\begin{eqnarray}
\label{packed}
S_{\text{packed}} = \mathcal{N}\left(\sum_i S_i \right),
\end{eqnarray}
where we sum and normalize saliency maps for all frames in one video. In Eq. (\ref{packed}), $\mathcal{N}$ is a normalization operator which normalizes the saliency map linearly to the dynamic range $[0, 1]$.
Besides, we extract several frames from the two videos and extract the local image patch in the frame for better visualization.
The corresponding saliency maps for the local images patch are also extracted.
From the results in local image patch, we can see that fixations are located on moving objects. Careful calibrations ensure that the head-eye movements are well captured by the eye tracker and motion sensors in the HMD.
To analyse how the virtual information change the distribution of visual attention, Fig. \ref{fig_show} also presents the comparisons of saliency map between the original video and the augmented video.
From the packed saliency map, we can observe the similarity between the distribution of visual attention for the original video and the video with augmentation information.
From the local image patch and the corresponding saliency map, we can observe that the augmentation can attract the visual attention to the object itself.

We also calculate the velocity of head movements. Fig. \ref{fig_velocity} shows the distribution of velocity of different directions and magnitudes, from which we can see that the eye movements for augmented videos are more concentrated and regular, and are more coherent with the movements of objects. So, the augmented virtual information will change the movements of eyes, and thus change the distribution of visual saliency distribution.
%
However, the relationship between the augmentations and the real-world perception is not distinguished in existing saliency prediction models.
To address the problem, we perform a large-scale model comparison in the experiments and present some results in advance here to show that the precarious prediction will occur if the augmented information is not distinguished.
From the results in Fig. \ref{fig_show} we can observe that the prediction results from SalGAN \cite{Pan_2017_SalGAN} for the ``bicycle training'' is consistent with the ground truth.
We think the complex background has the luminance and contrast masking effects.
However, the prediction results for the ``car race'' is more sensitive to the augmented contents. More predicted visual attentions are concentrated on the label, but the ground truth shows that the bounding box and label have limited influence on the visual attention distribution.

\subsection{Experimental Setup}
\textbf{Test Datasets.}
There are some panoramic video eye-tracking datasets in the literature, but they only include the recording of head and eye movements for panoramic videos.
To investigate the visual behavior of viewers when they are experiencing AR, we construct the ARVR dataset.
In the ARVR dataset, 12 panoramic videos and 12 annotated AR videos are included.
Head movements and eye movements are recorded for both panoramic videos and AR videos.
We refer to the videos with virtual augmentations as Subset I, and the original videos as Subset II.
We mainly use Subset I as the testbed.
Subset II is used as a complementary to measure the performance of our model in terms of saliency prediction for panoramic videos.

\begin{figure*}[!t]
\centering
\includegraphics[width=5.68in]{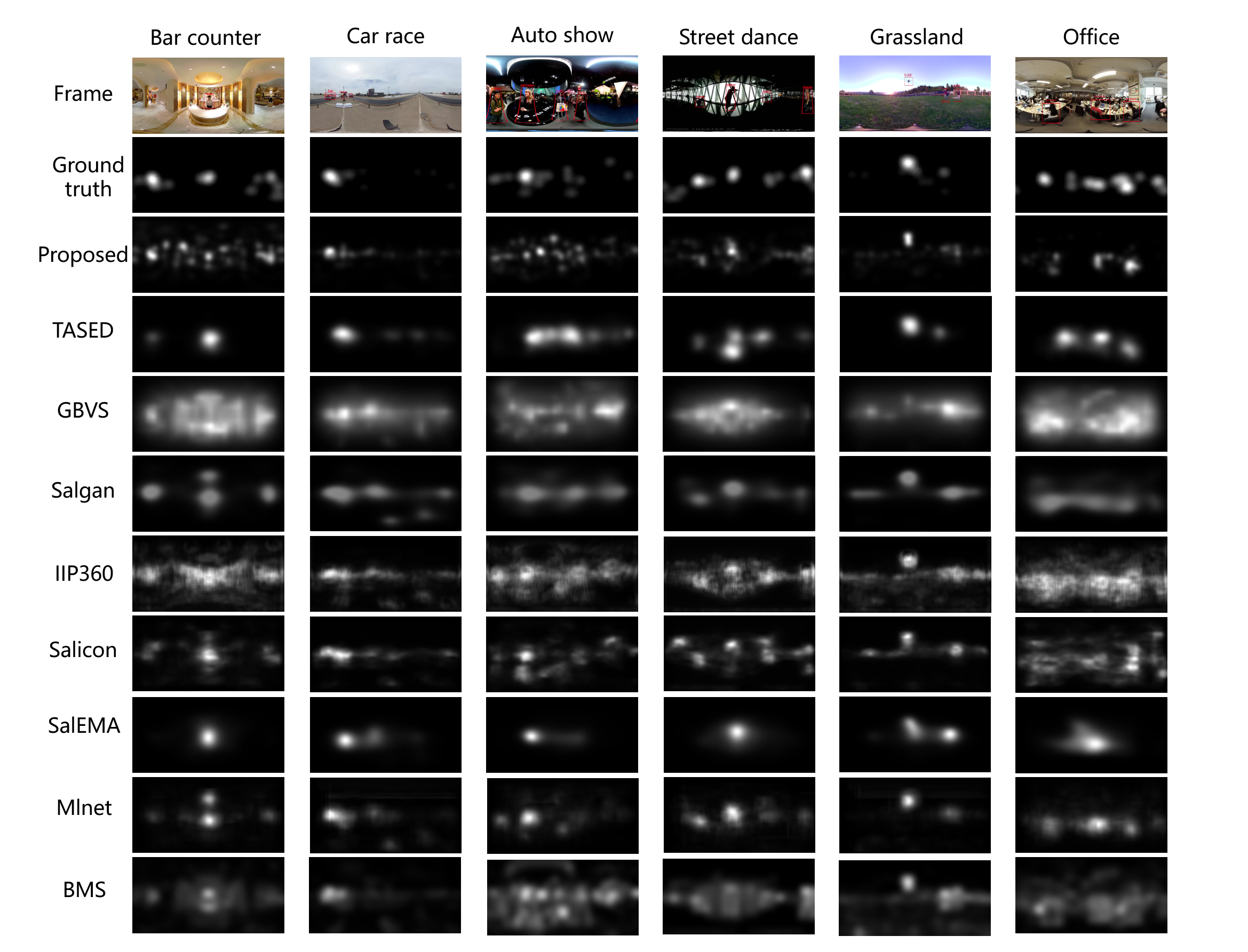}
\centering
\caption{Examples of saliency maps. Six example frames are from: `Bar counter', `Car race', `Auto show', `Street dance', `Grassland', and `Office'. The ground-truth saliency map, and the saliency predictions from our model, TASED, GBVS, Salgan, IIP360, Salicon, SalEMA, Mlnet, and BMS are shown. }
\label{modelresults}
\vspace{-0.4cm}
\end{figure*}

\begin{table}[!t]
\center
\scriptsize
\caption{ \textsc{Performance Comparison With State-of-the-Art Saliency Models on Subset I}. }
\vspace{0.2cm}
\label{table:all}
\begin{tabular}{|c|c|c|c|c|c|}
\hline
\multicolumn{2}{|c|}{\multirow{2}*{Category \& Method}} & \multicolumn{4}{|c|}{Performance} \\
\cline{3-6}
\multicolumn{2}{|c|}{~} & AUC-Judd $\uparrow$ & NSS $\uparrow$ & KL $\downarrow$ & CC $\uparrow$\\
\hline
 \multirow{3}*{$\begin{array}{c}\text{Non-deep} \\ \text{Static} \end{array}$} &BMS \cite{zhang2013saliency} &0.697 &1.608 &8.437 &0.393 \\ \cline{2-6}
 &GBVS \cite{harel2007graph} &0.667 &1.240 &9.361 &0.321 \\ \cline{2-6}
 &Itti \cite{itti1998model} &0.680 &1.358 &9.024 &0.348 \\ \hline
 \multirow{3}*{$\begin{array}{c}\text{Deep} \\ \text{Static} \end{array}$} &Mlnet \cite{cornia2016deep} &0.737 &2.046 &7.460 &0.468 \\ \cline{2-6}
 &Salgan \cite{Pan_2017_SalGAN} &0.736 &1.718 &7.695 &0.421 \\ \cline{2-6}
 &Salicon \cite{jiang2015salicon} &0.738 &1.775 &8.171 &0.386 \\ \hline
 \multirow{2}*{$\begin{array}{c}\text{Non-deep} \\ \text{Dynamic} \end{array}$} &PQFT \cite{guo2008spatio} &0.684 &1.756 &8.403 &0.303 \\ \cline{2-6}
 &SeR \cite{seo2009static} &0.676 &1.784 &8.363 &0.265 \\ \hline
 \multirow{2}* {$\begin{array}{c}\text{Deep} \\ \text{Dynamic} \end{array}$}&SalEMA \cite{linardos2019simple} &0.728 &1.812 &6.222 &0.429 \\ \cline{2-6}
 &TASED \cite{min2019tased} &0.743 &2.063 &5.475 &0.481 \\ \hline
 \multirow{4}*{$\begin{array}{c}\text{Panoramic} \\ \text{Static} \end{array}$} &SJTU\_H \cite{zhu2018prediction} &0.759 &1.559 &8.324 &0.420 \\ \cline{2-6}
 &SJTU\_HE \cite{zhu2018prediction} &0.759 &1.578 &8.304 &0.423 \\ \cline{2-6}
 &SalNet360 \cite{monroy2018salnet360} &0.733 &1.310 &9.566 &0.341 \\ \cline{2-6}
 &IIP360 \cite{zhang2018video} &0.762 &1.984 &8.171 &0.437 \\ \hline
 \multirow{6}*{Proposed} &$\text{VBAS}_{f_1,f_4,f_5}$ &0.801 &1.925 &5.082 &\textbf{0.553} \\ \cline{2-6}
 &$\text{VBAS}_{f_2,f_4,f_5}$ &0.773 &1.729 &5.690 &0.452 \\ \cline{2-6}
 &$\text{VBAS}_{f_3,f_4,f_5}$ &\textbf{0.834} &\textbf{2.387} &\textbf{4.870} &0.549 \\ \cline{2-6}
 &$\text{VBAS}_{f_3,f_4}$ &0.739 &1.924 &5.330 &0.439 \\ \cline{2-6}
 &$\text{VBAS}_{f_3,f_5}$ &0.793 &2.119 &5.113 &0.521 \\ \cline{2-6}
\hline
\end{tabular}
\vspace{-0.35cm}
\end{table}


\begin{table*}[!h]
\center
\scriptsize
\caption{ \textsc{Performance Comparison With State-of-the-Art Saliency Models on Subset I. AUC Values on 12 Videos are Shown. The Top 2 Models are in Bold, and the Top 1 Non-panoramic Model is in Red. } }
\vspace{0.2cm}
\label{table:auc}
\begin{tabular}{|c|c|c|c|c|c|c|c|c|c|c|c|c|c|}
\hline
\multicolumn{2}{|c|}{\multirow{2}*{Category \& Method}} & \multicolumn{3}{|c|}{Bar counter} & \multicolumn{3}{|c|}{Auto show} & \multicolumn{3}{|c|}{Underwater} & \multicolumn{3}{|c|}{Street dance} \\
\cline{3-14}
\multicolumn{2}{|c|}{~} & Mean& std & All & Mean& std & All& Mean& std & All& Mean& std & All \\
\hline
 \multirow{3}*{$\begin{array}{c}\text{Non-deep} \\ \text{Static} \end{array}$} &BMS \cite{zhang2013saliency} &0.718 &0.031 &0.704 &0.686 &0.024 &0.676 &0.682 &0.037 &0.671 &0.712 &0.031 &0.700\\ \cline{2-14}
 &GBVS \cite{harel2007graph} &0.703 &0.041 &0.691 &0.656 &0.030 &0.647 &0.638 &0.033 &0.633 &0.650 &0.048 &0.639\\ \cline{2-14}
 &Itti \cite{itti1998model} &0.698 &0.026 &0.684 &0.672 &0.027 &0.673 &0.659 &0.034 &0.655 &0.677 &0.037 &0.665\\ \hline
 \multirow{3}*{$\begin{array}{c}\text{Deep} \\ \text{Static} \end{array}$} &Mlnet \cite{cornia2016deep} &{\color{red}0.756} &0.041 &0.739 &{\color{red}0.759} &0.034 &{\color{red}0.744} &0.654 &0.057 &0.661 &0.763 &0.035 &0.745\\ \cline{2-14}
 &Salgan \cite{Pan_2017_SalGAN} &0.754 &0.037 &0.738 &0.737 &0.034 &0.721 &0.687 &0.050 &0.675 &0.746 &0.039 &0.729\\ \cline{2-14}
 &Salicon \cite{jiang2015salicon} &0.748 &0.033 &0.736 &0.750 &0.030 &0.743 &0.684 &0.048 &{\color{red}0.683} &0.757 &0.031 &0.741\\ \hline
 \multirow{2}*{$\begin{array}{c}\text{Non-deep} \\ \text{Dynamic} \end{array}$} &PQFT \cite{guo2008spatio} &0.698 &0.032 &0.691 &0.683 &0.029 &0.682 &0.624 &0.047 &0.629 &0.699 &0.029 &0.697\\ \cline{2-14}
 &SeR \cite{seo2009static} &0.666 &0.069 &0.701 &0.682 &0.031 &0.675 &0.555 &0.074 &0.571 &0.689 &0.052 &0.693\\ \hline
 \multirow{2}* {$\begin{array}{c}\text{Deep} \\ \text{Dynamic} \end{array}$}&SalEMA \cite{linardos2019simple} &0.732 &0.048 &0.717 &0.740 &0.035 &0.728 &{\color{red}0.694} &0.037 &0.676 &0.701 &0.040 &0.694\\ \cline{2-14}
 &TASED \cite{min2019tased} &0.752 &0.039 &{\color{red}0.747} &0.737 &0.033 &0.732 &0.686 &0.041 &0.668 &{\color{red}0.766} &0.025 &{\color{red}0.754}\\ \hline
 \multirow{4}*{$\begin{array}{c}\text{Panoramic} \\ \text{Static} \end{array}$} &SJTU\_H \cite{zhu2018prediction} &0.757 &0.033 &0.741 &0.765 &0.037 &0.749 &\textbf{0.736} &0.030 &\textbf{0.719} &0.752 &0.044 &0.738\\ \cline{2-14}
 &SJTU\_HE \cite{zhu2018prediction} &0.758 &0.033 &0.742 &0.764 &0.037 &0.749 &\textbf{0.735} &0.031 &0.718 &0.753 &0.043 &0.738\\ \cline{2-14}
 &SalNet360 \cite{monroy2018salnet360} &0.756 &0.031 &0.739 &0.746 &0.025 &0.735 &0.684 &0.034 &0.682 &0.709 &0.034 &0.700\\ \cline{2-14}
 &IIP360 \cite{zhang2018video} &\textbf{0.764} &0.027 &\textbf{0.748} &\textbf{0.767} &0.029 &\textbf{0.750} &0.704 &0.050 &0.697 &\textbf{0.783} &0.028 &\textbf{0.766}\\ \hline
 \multirow{1}*{Proposed} &VBAS &\textbf{0.837} &0.038 &\textbf{0.805} &\textbf{0.838} &0.031 &\textbf{0.815} &0.726 &0.059 &\textbf{0.737} &\textbf{0.851} &0.035 &\textbf{0.814}\\
\hline \hline
\multicolumn{2}{|c|}{\multirow{2}*{Category \& Method}} & \multicolumn{3}{|c|}{Lawn} & \multicolumn{3}{|c|}{Airport} & \multicolumn{3}{|c|}{Car race} & \multicolumn{3}{|c|}{Spacesuit Training} \\
\cline{3-14}
\multicolumn{2}{|c|}{~} & Mean& std & All & Mean& std & All& Mean& std & All& Mean& std & All \\
\hline
 \multirow{3}*{$\begin{array}{c}\text{Non-deep} \\ \text{Static} \end{array}$} &BMS \cite{zhang2013saliency} &0.712 &0.032 &0.697 &0.700 &0.033 &0.685 &0.733 &0.030 &0.710 &0.624 &0.040 &0.625\\ \cline{2-14}
 &GBVS \cite{harel2007graph} &0.695 &0.042 &0.677 &0.684 &0.038 &0.668 &0.717 &0.042 &0.699 &0.604 &0.043 &0.603\\ \cline{2-14}
 &Itti \cite{itti1998model} &0.706 &0.034 &0.692 &0.690 &0.030 &0.677 &0.723 &0.030 &0.704 &0.619 &0.039 &0.616\\ \hline
 \multirow{3}*{$\begin{array}{c}\text{Deep} \\ \text{Static} \end{array}$} &Mlnet \cite{cornia2016deep} &0.740 &0.057 &0.716 &0.736 &0.036 &0.722 &0.769 &0.043 &0.746 &0.688 &0.062 &0.676\\ \cline{2-14}
 &Salgan \cite{Pan_2017_SalGAN} &0.741 &0.064 &0.724 &0.723 &0.032 &0.707 &0.788 &0.031 &0.767 &0.685 &0.061 &0.673\\ \cline{2-14}
 &Salicon \cite{jiang2015salicon} &{\color{red}0.769} &0.041 &{\color{red}0.750} &{\color{red}0.747} &0.027 &{\color{red}0.734} &{\color{red}0.798} &0.028 &{\color{red}0.778} &0.686 &0.057 &0.685\\ \hline
 \multirow{2}*{$\begin{array}{c}\text{Non-deep} \\ \text{Dynamic} \end{array}$} &PQFT \cite{guo2008spatio} &0.700 &0.032 &0.698 &0.681 &0.046 &0.683 &0.724 &0.054 &0.727 &0.621 &0.052 &0.640\\ \cline{2-14}
 &SeR \cite{seo2009static} &0.697 &0.055 &0.698 &0.687 &0.053 &0.688 &0.736 &0.030 &0.710 &0.645 &0.047 &0.649\\ \hline
 \multirow{2}* {$\begin{array}{c}\text{Deep} \\ \text{Dynamic} \end{array}$}&SalEMA \cite{linardos2019simple} &0.714 &0.034 &0.695 &0.710 &0.041 &0.697 &0.759 &0.048 &0.735 &0.670 &0.053 &0.660\\ \cline{2-14}
 &TASED \cite{min2019tased} &0.742 &0.028 &0.725 &0.739 &0.042 &0.728 &0.784 &0.044 &0.763 &{\color{red}0.707} &0.059 &{\color{red}0.690}\\ \hline
 \multirow{4}*{$\begin{array}{c}\text{Panoramic} \\ \text{Static} \end{array}$} &SJTU\_H  \cite{zhu2018prediction} &0.785 &0.050 &0.765 &0.743 &0.048 &0.722 &0.817 &0.039 &0.801 &\textbf{0.732} &0.052 &\textbf{0.719}\\ \cline{2-14}
 &SJTU\_HE  \cite{zhu2018prediction} &\textbf{0.785} &0.050 &\textbf{0.765} &0.744 &0.047 &0.724 &\textbf{0.818} &0.040 &\textbf{0.801} &0.731 &0.052 &0.719\\ \cline{2-14}
 &SalNet360 \cite{monroy2018salnet360} &0.769 &0.034 &0.752 &0.660 &0.047 &0.686 &0.785 &0.029 &0.766 &0.669 &0.056 &0.682\\ \cline{2-14}
 &IIP360 \cite{zhang2018video} &0.782 &0.034 &0.764 &\textbf{0.763} &0.030 &\textbf{0.740} &0.799 &0.027 &0.782 &0.723 &0.045 &0.711\\ \hline
 \multirow{1}*{Proposed} &VBAS &\textbf{0.846} &0.050 &\textbf{0.821} &\textbf{0.841} &0.036 &\textbf{0.807} &\textbf{0.897} &0.025 &\textbf{0.858} &\textbf{0.789} &0.053 &\textbf{0.775}\\
\hline \hline
\multicolumn{2}{|c|}{\multirow{2}*{Category \& Method}} & \multicolumn{3}{|c|}{Grassland} & \multicolumn{3}{|c|}{Bicycle train} & \multicolumn{3}{|c|}{Office} & \multicolumn{3}{|c|}{Time square} \\
\cline{3-14}
\multicolumn{2}{|c|}{~} & Mean& std & All & Mean& std & All& Mean& std & All& Mean& std & All \\
\hline
 \multirow{3}*{$\begin{array}{c}\text{Non-deep} \\ \text{Static} \end{array}$} &BMS \cite{zhang2013saliency} &0.735 &0.028 &0.716 &0.742 &0.024 &0.738 &0.709 &0.020 &0.702 &0.615 &0.037 &0.614\\ \cline{2-14}
 &GBVS \cite{harel2007graph} &0.692 &0.029 &0.672 &0.710 &0.032 &0.696 &0.677 &0.025 &0.666 &0.582 &0.037 &0.583\\ \cline{2-14}
 &Itti \cite{itti1998model} &0.710 &0.027 &0.690 &0.708 &0.030 &0.700 &0.708 &0.020 &0.702 &0.594 &0.032 &0.598\\ \hline
 \multirow{3}*{$\begin{array}{c}\text{Deep} \\ \text{Static} \end{array}$} &Mlnet \cite{cornia2016deep} &0.774 &0.042 &0.753 &0.787 &0.033 &0.771 &{\color{red}0.773} &0.024 &{\color{red}0.763} &0.649 &0.052 &0.645\\ \cline{2-14}
 &Salgan \cite{Pan_2017_SalGAN} &0.782 &0.043 &0.760 &0.778 &0.031 &0.767 &0.766 &0.028 &0.752 &0.642 &0.043 &0.641\\ \cline{2-14}
 &Salicon \cite{jiang2015salicon} &0.782 &0.030 &0.761 &{\color{red}0.792} &0.028 &{\color{red}0.785} &0.739 &0.026 &0.735 &0.604 &0.055 &0.606\\ \hline
 \multirow{2}*{$\begin{array}{c}\text{Non-deep} \\ \text{Dynamic} \end{array}$} &PQFT \cite{guo2008spatio} &0.718 &0.037 &0.716 &0.720 &0.031 &0.728 &0.685 &0.031 &0.685 &0.651 &0.030 &0.643\\ \cline{2-14}
 &SeR \cite{seo2009static} &0.706 &0.061 &0.715 &0.728 &0.032 &0.719 &0.697 &0.033 &0.699 &0.629 &0.033 &0.601\\ \hline
 \multirow{2}* {$\begin{array}{c}\text{Deep} \\ \text{Dynamic} \end{array}$}&SalEMA \cite{linardos2019simple} &{\color{red}\textbf{0.792}} &0.027 &{\color{red}\textbf{0.771}} &0.777 &0.029 &0.763 &0.748 &0.030 &0.738 &{\color{red}0.696} &0.046 &{\color{red}0.688}\\ \cline{2-14}
 &TASED \cite{min2019tased} &0.782 &0.034 &0.754 &0.788 &0.029 &0.775 &0.751 &0.028 &0.747 &0.680 &0.038 &0.674\\ \hline
 \multirow{4}*{$\begin{array}{c}\text{Panoramic} \\ \text{Static} \end{array}$} &SJTU\_H  \cite{zhu2018prediction} &0.788 &0.045 &0.762 &0.791 &0.032 &0.778 &0.713 &0.032 &0.703 &0.727 &0.033 &0.715\\ \cline{2-14}
 &SJTU\_HE  \cite{zhu2018prediction} &0.788 &0.044 &0.762 &0.789 &0.031 &0.776 &0.713 &0.032 &0.703 &0.727 &0.033 &0.715\\ \cline{2-14}
 &SalNet360 \cite{monroy2018salnet360} &0.769 &0.045 &0.754 &0.783 &0.026 &0.773 &0.743 &0.024 &0.729 &\textbf{0.729} &0.031 &\textbf{0.723}\\ \cline{2-14}
 &IIP360 \cite{zhang2018video} &0.789 &0.027 &0.768 &\textbf{0.813} &0.018 &\textbf{0.803} &\textbf{0.778} &0.021 &\textbf{0.766} &0.683 &0.040 &0.679\\ \hline
 \multirow{1}*{Proposed} &VBAS &\textbf{0.886} &0.030 &\textbf{0.848} &\textbf{0.898} &0.024 &\textbf{0.863} &\textbf{0.832} &0.030 &\textbf{0.805} &\textbf{0.776} &0.034 &\textbf{0.751}\\
\hline \hline
\end{tabular}
\vspace{-0.3cm}
\end{table*}

\textbf{Competing Models.}
Some visual saliency models are compared with the proposed VBAS model.
Although some of these saliency models are designed for traditional 2D images and videos, they can still represent the state-of-the-art of fixation prediction.
Specifically, a total of 14 visual saliency models are selected as competitors: BMS \cite{zhang2013saliency}, GBVS \cite{harel2007graph}, Itti \cite{itti1998model}, Mlnet \cite{cornia2016deep}, Salgan \cite{Pan_2017_SalGAN}, Salicon \cite{jiang2015salicon}, PQFT \cite{guo2008spatio}, SeR \cite{seo2009static}, SalEMA \cite{linardos2019simple}, TASED \cite{min2019tased}, SJTU\_H \cite{zhu2018prediction}, SJTU\_HE \cite{zhu2018prediction}, SalNet360 \cite{monroy2018salnet360} and IIP360 \cite{zhang2018video}.
BMS, GBVS and Itti are traditional non-deep learning based saliency models for static scenes.
Mlnet, Salgan and Salicon are deep learning based saliency models for static scenes.
PQFT and SeR are traditional non-deep learning based video saliency models for dynamic scenes.
SalEMA and TASED are deep learning based video saliency models for dynamic scenes.
SJTU\_H, SJTU\_HE, SalNet360 and IIP360 are saliency models for panoramas.

\textbf{Evaluation Criteria.}
Several criteria are available for comparing the saliency models.
We employ the toolbox to measure the performance of saliency prediction on 360 degree contents \cite{gutierrez2018toolbox}.
We use 4 saliency model evaluation metrics, wherein
\textbf{AUC-Judd} \cite{judd2012benchmark} calculates the area under the receiver operating characteristic (ROC) curve. To create the curve, the threshold value is swept in the range of saliency values at fixation locations. A value of 1 indicates a perfect classification for AUC-Judd.
Normalized Scanpath Saliency (\textbf{NNS}) \cite{Peters2005Components} metric evaluates the prediction accuracy of models by calculating the mean value of the predicted saliency at the ground-truth fixations.
A higher NSS value indicates a better prediction result.
Kullback-Leibler (\textbf{KL}) \cite{tatler2005visual} divergence measures the distance between the distribution of predicted saliency map and the distribution of ground-truth map.
If the prediction exactly matches with the ground truth, the KL value equals zero. The KL value varies from zero to infinity.
Correlation coefficient (\textbf{CC}) \cite{jost2005assessing} evaluates the linear correlation between the prediction and the ground truth.
And a value of 1 indicates a perfect correlation.

\textbf{Implementation Details.}
The proposed method involves the spatial and temporal saliency estimation.
The designed CNN-based network is employed to predict spatial saliency, which is trained on the LSUN dataset \cite{LSUN}.
The FlowNet \cite{dosovitskiy2015flownet} is employed to predict the motion information as the temporal saliency.
All videos are analyzed at a reduced frame rate.
We sample every five frames and measure the consistency between predictions on the frames and the ground truth.
The mean performance of the frames are reported.

\subsection{Performance Evaluations and Comparisons}
We first conduct the ablation study and test all the saliency models on Subset I. The results are summarized in Table \ref{table:all}.
The performance is first averaged over frames of the same video and then averaged over all videos.
We give the intuitive illustration of the prediction results on different videos in Fig. \ref{modelresults}.
%
Performance of models on each stimulus is presented in Table \ref{table:auc}.
For each stimulus, the first column reports the averaged result over all frames of the video.
The second column reports the standard deviation over all frames.
For each stimulus, we sum and normalize the saliency over all frames, and form the packed saliency map that is defined in Eq. (\ref{packed}). The computation results on packed saliency is reported in the third column.
Besides, we test all the saliency models on Subset II. The results are summarized in Table \ref{table:allori}.

\textbf{Ablation Study.}
In our model, we employ different feature fusion strategies ($f_1, f_2, f_3$).
Besides, we consider the impact of augmentations according to augmentation types ($f_4$).
And we combine the extracted visual cues with the characteristics of viewing behaviors to define edge weights on graphs which are interpreted as Markov chains ($f_5$).
We conduct the ablation study to single out the contribution of each component.
The experimental results are summarized in Table \ref{table:all}.
From the results we can make the observation that the proposed model can be greatly improved by $f_5$, which can ensure our model concentrate on significant block maps rather than uniform weights.
Besides, an improvement is made by $f_4$, which consider the interplay between virtual information and reality to make a distinction between augmentation types.
We also test the performance of the proposed method using different fusion strategies.
In our experiments, summation after normalization ($f_3$) performs better than product after normalization ($f_1$) and max after normalization ($f_2$).

\textbf{Qualitative Analysis.}
The performance of the proposed model has been explored in the qualitative perspective.
Videos can be characterized by content complexity.
The spatial and temporal complexity contribute to the content diversity.
Some videos contain obvious foreground objects, while others have no obvious contrast between background and foreground.
Besides, some videos contain rich texture information, while others contain the plain areas of less texture information.
And some videos have rich and distributed motion information, while others have concentrated motion.
We choose 6 videos and extract the example frames.
The ``Bar counter" is indoor scene and contains concentrated motion and salient objects, and rich texture information in the background.
The ``Car race" is outdoor scene and contains concentrated motion and salient objects.
The ``Auto show" has no obvious contrast between background and foreground and have multiple salient objects.
The ``Street dance" has no background motion and several salient objects.
The ``Grassland" has some salient objects that are away from the equator.
The ``Office" has complex low-level features.

From the results we can observe that our computation in the block image can result in the much fine-grained motion predictions and saliency predictions.
The prediction results from TASED and SalEMA show that they can highlight the motion area, but they do not pay much attention on some static yet salient areas.
Besides, some complex textures do not attract attentions, but will interfere the predictions of some models like GBVS and BMS, and make the prediction too uniformly distributed and uninformative.
From the ``Bar counter" we can see that our model's prediction can omit the complex texture in the background and meanwhile capture the salient areas.
From the ``Car race", we can see that our model can select the prominent object near the equator rather than highlight the whole area that is near the equator.
And on the ``Grassland", we can see that our model can also highlight the salient object that is away from the equator.
The complex background in ``Auto show" and ``Office" is proven to have limited impact on our model.
It can also be observed in ``Street dance" that our model can focus more on the object itself rather than the virtual information.
Because our model can adapt the weighting strategy according to the information amount and the type of the augmentations.

\begin{table}[!tb]
\center
\scriptsize
\caption{ \textsc{Performance Comparison With State-of-the-Art Saliency Models on Subset II.} }
\vspace{0.2cm}
\label{table:allori}
\begin{tabular}{|c|c|c|c|c|c|}
\hline
\multicolumn{2}{|c|}{\multirow{2}*{Category \& Method}} & \multicolumn{4}{|c|}{Performance} \\
\cline{3-6}
\multicolumn{2}{|c|}{~} & AUC-Judd $\uparrow$ & NSS $\uparrow$ & KL $\downarrow$ & CC $\uparrow$\\
\hline
 \multirow{3}*{$\begin{array}{c}\text{Non-deep} \\ \text{Static} \end{array}$} &BMS \cite{zhang2013saliency} &0.668 &1.587 &8.328 &0.390\\ \cline{2-6}
 &GBVS \cite{harel2007graph} &0.679 &1.334 &8.638 &0.378\\ \cline{2-6}
 &Itti \cite{itti1998model} &0.707 &1.340 &9.002 &0.322\\ \hline
 \multirow{3}*{$\begin{array}{c}\text{Deep} \\ \text{Static} \end{array}$} &Mlnet \cite{cornia2016deep} &0.754 &2.068 &7.749 &0.484\\ \cline{2-6}
 &Salgan \cite{Pan_2017_SalGAN} &0.713 &1.811 &7.737 &0.399\\ \cline{2-6}
 &Salicon \cite{jiang2015salicon} &0.726 &1.775 &8.131 &0.365\\ \hline
 \multirow{2}*{$\begin{array}{c}\text{Non-deep} \\ \text{Dynamic} \end{array}$} &PQFT \cite{guo2008spatio} &0.685 &1.800 &7.597 &0.340\\ \cline{2-6}
 &SeR \cite{seo2009static} &0.730 &1.764 &8.042 &0.320\\ \hline
 \multirow{2}* {$\begin{array}{c}\text{Deep} \\ \text{Dynamic} \end{array}$}&SalEMA \cite{linardos2019simple} &0.764 &1.906 &6.074 &0.474\\ \cline{2-6}
 &TASED \cite{min2019tased} &0.794 &2.123 &5.445 &\textbf{0.522}\\ \hline
 \multirow{4}*{$\begin{array}{c}\text{Panoramic} \\ \text{Static} \end{array}$} &SJTU\_H \cite{zhu2018prediction} &0.748 &1.627 &7.503 &0.417\\ \cline{2-6}
 &SJTU\_HE \cite{zhu2018prediction} &0.749 &1.634 &7.464 &0.420\\ \cline{2-6}
 &SalNet360 \cite{monroy2018salnet360}&0.712 &1.387 &9.956 &0.339\\ \cline{2-6}
 &IIP360 \cite{zhang2018video} &0.743 &2.010 &7.526 &0.481\\ \hline
 \multirow{1}*{Proposed} &$\text{VBAS}$ &\textbf{0.807} &\textbf{2.155} &\textbf{5.188} &0.521\\
\hline
\end{tabular}
\vspace{-0.5cm}
\end{table}

\textbf{Comparison with Current Models.}
We test all the saliency models on Subset I, and average all results over frames of the video.
From Table \ref{table:all}, it is observed that the proposed VBAS model can be comparable to the state-of-the-art saliency models.
For non-deep methods, the dynamic methods do not show superior performance compared with static methods.
For deep-learning-based methods, the dynamic methods show significant superior performance in terms of KL compared with static methods, but the dynamic methods do not show superior performance in AUC-Judd, NSS and CC.
The results are consistent with some recent study \cite{wang2018revisiting}.
This phenomenon may be caused by insufficient study of the visual behavior and visual attention mechanism for 360-degree dynamic scenes.

AUC-Judd of models on each stimulus is presented in Table \ref{table:auc}.
Compared with other saliency models, the proposed model achieves good performance and generalizability on the videos.
Besides, the IIP360, Salicon and TASED also have good performance.
%
From the second column in Table \ref{table:auc}, which reports the standard deviation over all frames, we can observe that for the proposed model, there is small dispersion of the prediction accuracy over frames.
For each stimulus, we form the packed saliency map that is defined in Eq. (\ref{packed}).
From the results in the third column in Table \ref{table:auc}, it can be observed that the AUC-Judd result on packed saliency is consistent with the result that is averaged on each frame.

Although the proposed model is designed for augmented panoramic contents, we also test the model in general panoramic scenes to test its generalizability.
The same competing models are evaluated on Subset II.
The performance results are summarized in Table \ref{table:allori}.
It can be observed that the proposed model delivers the most promising result. Besides, TASED also perform well.
Although the superiority of the proposed model on Subset II is not so significant like on Subset I, it can also achieve the best performance in AUC-Judd, NSS and KL.
The results on Subset II show that our model that involves computation on the local blocks and the estimation of fraction of the visual attention for each block image can be a promising way to predict the saliency for 360-degree contents.

Admitting that AR of video-through and see-through types can offer users interactions with the real world, and we only investigate the impact of augmentations, how to properly deal with interactions with the real world will be part of our future work.
In real practice, there are many kinds of augmentations, e.g., static augmentation, dynamic augmentation, marker based augmentation and computer-generated imagery.
Enriching the dataset by simulating different kinds of augmentations and investigating visual saliency distribution on these stimuli will be another possible future work.

\vspace{-0.2cm}
\section{Conclusion}
We construct an ARVR saliency dataset.
The saliency annotations of head and eye movements for both original and augmented videos are collected, which together constitute the ARVR dataset.
%
We also design a model which is capable of solving the saliency prediction problem in AR.
We extract local block images to simulate the viewport and offset the projection distortion.
Conspicuous visual cues in local viewports are extracted in the spatial domain and temporal domain.
We estimate the joint effects of adversarial augmentation and complementary augmentation and apply the spatial weighting strategy accordingly.
We build graphs on block images which are interpreted as Markov chains.
In the computation of edge weights, we emphasize the characteristics of viewing behaviors and visual saliency mechanism.
The fraction of the visual attention that is diverted to each block image is estimated through equilibrium distribution of this chain.
The experimental results demonstrate the effectiveness of our method.


\bibliographystyle{IEEEtran}
\bibliography{arsaliency}

\end{document}